\documentstyle[prb,amssymb,twocolumn,aps]{revtex}

\begin{document}
\tolerance 50000
%%%%\preprint{
%%%%\begin{minipage}[t]{1.8in}
%%%%\hfill LPQTH-94/?
%%%%\end{minipage}
%%%%}

\draft

\title{Phase diagram of the two-dimensional t--J model at low doping}

\author{D. Poilblanc}

\address{
Universit\'e Paul Sabatier \\
Laboratoire de Physique Quantique \\
118 route de Narbonne, 31062 Toulouse, France \\
}

\twocolumn[
\date{Received}
\maketitle
\widetext

\vspace*{-1.0truecm}

\begin{abstract}
\begin{center}
\parbox{14cm}{
The phase diagram of the
planar t--J model at small hole doping is investigated
by finite size scaling of exact diagonalisation data
of $\sqrt{N}\times \sqrt{N}$ clusters ($N\leq 26$).
Hole-droplet binding energies, compressibility and static
spin and charge correlations are calculated.
Short range antiferromagnetic correlations can produce
attractive forces between holes leading to a very rich phase diagram
including a liquid of d-wave hole pairs (for $J/t\gtrsim 0.2$),
a liquid of hole droplets (quartets) for larger J/t ratios ($J/t\gtrsim 0.5$)
and, at even larger coupling J/t, an instability towards phase separation.
}
\end{center}
\end{abstract}

\pacs{
\hspace{1.9cm}
}
]
\narrowtext
Studying the behavior of holes in two-dimensionnal (2D) antiferromagnets
is crutial to understand the origin of pairing in the high-$T_c$
cuprate superconductors. In these materials
chemical substitutions in the parent stochiometric compound lead to
injecting mobile holes in the $CuO_2$ antiferromagnetic planes.
Besides transport properties these holes will also drastically
affect the antiferromagnetic (AF) correlations in the planes.
On the theoretical side, motion of holes in antiferromagnets
can be simply described by the so-called t--J
model \cite{anderson,zhang_rice}, a strong coupling version of the
well-known Hubbard model. Previous numerical studies
\cite{reviews} have given reliable informations, specially in the
limit of a single hole. It is believed that holes behave
like quasiparticles, at least at the bottom of the coherent band,
although spin fluctuations strongly
enhance their effective masses
and reduce their quasiparticle weights
\cite{single_hole}.

Finite size scaling analysis becomes easier at commensurate
densities such as $n=1/2$ (quarter filling). In this case, exact
diagonalisations (ED) studies of the t--J model \cite{elbio_quarter}
have suggested the existence of superconducting correlations in the vicinity
of the phase separation phase \cite{emery_phase_separation}.
This regime is however quite far from the experimental situation.

At small but finite doping (eg electron density $n\sim 0.8$--$0.9$)
fewer theoretical results are known. However it is
believed that this class of models reproduces
successfully \cite{horsch_photo}
the large Fermi surface observed in angular resolved photoemission
studies in the metallic phase of doped high-$T_c$ materials.
%\cite{photo_large}.
In addition, possible observation of shadow bands due to
strong short range antiferromagnetic correlations has been suggested in both
experimental\cite{aebi} or theoretical studies\cite{adriana_shadow}.

The magnetic coupling J can generate
an effective coupling between holes.
This is particularly clear in the (unphysical) large
J/t regime where the magnetic energy cost is minimized by having
holes sitting on nearest neighbor sites. In this regime, the uniform state
becomes in fact unstable towards a phase separated state
\cite{emery_phase_separation}. High temperature expansions \cite{hightemp}
also predict phase separation for $J/t\gtrsim 1$.
However, small cluster calculations have shown that, for smaller
and more realistic J/t ratios individual pairs could be
stable \cite{binding,hole_hole}. Preliminary results
\cite{didier_wuerzburg} state that larger clusters of holes could
also form in the intermediate parameter range.
Other possible canditates in this parameter regime are non-uniform
striped phase \cite{didier_rice,zotos_clustering}.

In this letter, more insights into the nature of the
phase diagram at {\it small doping} are obtained from a detailed
numerical study.
Indeed, since analytic perturbation treatments are poorly controled in the
relevant physical regime,
exact diagonalizations of small 2D square clusters by the Lanczos
algorithm \cite{lanczos} were performed \cite{note1}.
Studies in the regime of small finite hole densities
are delicate since only different discrete values of the densities can
be achieved on different clusters and interpolations between
them become then necessary.
First, finite size scaling of binding energies of n-hole clusters
provides indication on the stability of liquids of pairs or droplets
in the {\it vanishing} hole concentration regime. Hole-hole correlations
obtained for $n\sim 0.85$ also confirms the stability of pairs
at finite density even at small J/t ratios.
The compressibility for arbitrary hole densities ($n\lesssim 0.8$)
and various system
sizes is calculated to perform an extrapolation to the thermodynamic
limit. The domain of the phase separated region is then
estimated. We also discuss the behavior of the static spin and charge
structure factors at intermediate density $n\sim 0.85$.

The t--J model defined on a square lattice reads,
\begin{displaymath}
H=t\sum_{{\bf x, y}\ N.N.}
\, c_{{\bf x},\sigma}^\dagger c_{{\bf y},\sigma}
+\frac{J}{2}\sum_{{\bf x, y}\ N.N.}({\bf S}_{\bf x}\cdot {\bf S}_{\bf y}
-\frac{n_{\bf x}n_{\bf y}}{4}),
%\label{ham}
\end{displaymath}
where $c_{{\bf x},\sigma}^\dagger$ and $\bf S_{\bf x}$
are creation and spin operators at site $\bf x$.
Ground state energies (GS) and equal-time correlation functions in the GS
are obtained on small $\sqrt{N}\times\sqrt{N}$ N--site clusters at
low hole densities by the Lanczos method. Typically N=18, 20 and 26.

Let us first consider a fixed finite number of holes $N_h=n$ ($N_h=2$, $4$)
on various clusters of increasing sizes.
These holes will form a n-particle bound state if the
binding energy
$\Delta_n=E_{h,n}+E_{h,0}-2E_{h,n/2}$, where $E_{h,n}$ is the GS
total energy for a system with $N_h=n$, converges towards a negative value
in the limit of infinite system size.
Strictly speaking such quantities give indications about the stability of
n-particle boundstates only in the limit of {\it vanishing} hole density.

A simple broken-bond counting argument shows that,
at large J/t, two holes injected in the AF can minimize the
local magnetic energy by forming a bound state.
Finite size scaling have shown \cite{binding} that this picture is actually
correct even down to $J=J_{B,2}\simeq 0.2$.
$\Delta_2$ calculated for N=26 is shown in Fig.\ \ref{f1} and becomes
negative when the paired state is stable.
The hole-hole pair has a $d_{x^2-y^2}$ orbital symmetry.
Fig.\ \ref{f2}(a) shows the hole-hole density correlations of the pair
$C_h({\bf r})=\frac{N}{N_h(N_h-1)}\big< n_h({\bf r}) n_h(0)\big>$
for all possible distances (${\bf r}\ne 0$) between the holes
compatible with the cluster shape.
Note that the normalization factor is chosen so that
$\sum_{\bf r\ne 0}C_h({\bf r})=1$.
Correlations at the intermediate distance of $\sqrt 2$
ie when the holes stay across the diagonal of a plaquette
on the {\it same sublattice} are dominant \cite{hole_hole}.
This singlet hole pair can in fact be viewed as a combination of a
triplet pair with a nearby spin triplet excitation \cite{hole_hole}.
The two-hole GS exhibits flux quantization in a ring
in units of $hc/2e$ \cite{twist}, also a signature of a paired state.
Lastly, pair formation is also consistent with
the observation that the dynamical response of the
d-wave pair creation operator exhibits on small clusters a
sharp $\delta$-peak of weight $Z_{2h}\propto J/t$ down to small
values of J/t \cite{binding}.

We now consider the possibility of larger droplets of holes.
Indeed, above $J_{B,2}$ residual interactions between pairs
might be sufficiently attractive to stabilize e.g. hole quartets.
GS energies of 4-holes
can be calculated for three possible orbital symmetries (s, p or d-wave)
of the wave function (with zero total momentum).
The lowest energy is obtained in the s-wave channel.
In order to estimate the onset of clustering
$\Delta_4$ has been evaluated on
various lattices of size up to N=26 and the data are displayed in
Fig.\ \ref{f1}.
Note that the critical value $J_{B,4}$
at which $\Delta_4$ changes sign depends weakly on the system size
while the slope $|\partial \Delta_4/\partial J|$ at this
point decreases for increasing size.

Fig.\ \ref{f1} then strongly suggests that
when $J/t\in [J/t|_{B,2},J/t|_{B,4}]$ with $J/t|_{B,2}\sim 0.2$ and
$J/t|_{B,4}\sim 0.5$ individual hole pairs exist without forming larger
clusters. This results based on a scaling of $\Delta_n$ is, strictly
speaking, only valid in the limit of vanishing hole density.
In order to investigate the stability of the pairs at finite doping
we have calculated the hole-hole correlations $C_h({\bf r})$ in a
26-site cluster at density $n\sim 0.85$ (i.e. with 4 holes). The results shown
in
Fig.\ \ref{f2}(b) for various separations $|{\bf r}|$ reveal dominant
correlations at distance $\sqrt{2}$ as in the case of a single pair (see
Fig.\ \ref{f2}(a)).
Note also that the density correlations at the largest
distances available in the cluster remain always significant
in this parameter regime which is consistent with the existence of separate
pairs in the cluster.

Formation of droplets should not be confused with phase separation (PS)
between hole-free and hole-rich phases even though both phenomena
have the same microscopic origin.
The issue of phase separation can be addressed by studying the inverse
compressibility $\kappa^{-1}$ defined by
$\kappa^{-1}=\partial^2 (E_{h,n}/N)/\partial n_h^2$.
In a uniform system $\kappa^{-1}$ is finite and positive.
On finite clusters $\kappa^{-1}<0$ signals
the instability of the homogeneous phase. However, we stress that a
rigourous determination of the PS region can only be achieved by
a finite size scaling at {\it constant hole density}. Such a study is
attempted in Figs.\ \ref{f3}(a)(b) showing the GS energy per site vs
hole density $n_h$ for two different system sizes. The thermodynamic limit is
obtained in two steps; (i) an interpolation between data points at constant
cluster size and (ii) an extrapolation at constant hole density assuming
$N^{-3/2}$ finite size corrections. We have checked by considering
other system sizes (data not shown for clarity) that such a $N^{-3/2}$
behavior is actually very well satisfied provided that $J\gtrsim n_h t$.
For $J\gtrsim 1$ it is clear that the PS region extends from $n_h=0$ to
$n_h\simeq 0.12$. For smaller values of J $|\kappa^{-1}|$ becomes
quickly very small and we can assume that there is no sign of PS for
$J/t<0.75$ \cite{note} consistently with the results
obtained by high-temperature expansions \cite{hightemp}.
Our results then suggest that a liquid of hole quartets are stable in a
small region of the phase diagram, as a precursor of the PS instability line.
This should be contrasted to the small {\it electron} density case
where the gas of electron quartets is never stable \cite{low_density}.

We finish this study by the investigation of the static spin and charge
structure factors defined by
$S({\bf q})=\frac{1}{N}\sum_{\bf r}
\big< S_z({\bf r}) S_z(0) \big> e^{i{\bf q}\cdot{\bf r}}$
and $N({\bf q})=\frac{1}{N}\sum_{\bf r}
(\big< n_h({\bf r}) n_h(0) \big> - n_h^2)
e^{i{\bf q}\cdot{\bf r}}$ respectively.
The data obtained for N=26 with a density of
$n\sim 0.85$ (four holes) are shown in Figs.\ \ref{f4}(a) and (b).
A smooth interpolation between the discrete ${\bf q}$-points of the reciprocal
lattice of the 26-site cluster has been performed assuming that the
correlations in real space remain small at distances larger than
the cluster size. $S({\bf q})$ in Fig.\ \ref{f4}(a) shows a pronounced
peak at $(\pi,\pi)$
even for small values of J. This indicates that large anti-ferromagnetic
spin correlations ($\xi_{AF}\sim 3$) still survive for hole doping as large
as $15\%$. $N({\bf q})$ shown in Fig.\ \ref{f4}(b) exhibits
along the $\Gamma$-M line a behavior very similar to
non-interacting spinless fermions with nearest-neighbor hopping
(dotted line). However, a clear dip is observed at X. This
behavior cannot be explained by a simple Fermi surface effect. For example,
a different tight-binding spinless model whose dispersion has a minimum
at $\Sigma$ (momentum $(\pi/2,\pi/2)$) would give much more structure
than observed (other dotted line). We interpret the dip at X as the signature
of strong short-range correlations between holes characteristic of the paired
state.

We conclude this paper by suggesting a possible phase diagram in
Fig.\ \ref{f5} based on the results discussed above. When J exceeds some
critical values $J_{B,2}$ and $J_{B,4}$ holes injected into
the antiferromagnetic phase form a liquid of d-wave pairs and a liquid of
quartets (D) respectively. At larger J/t ratios the t--J model phase separates
(PS). Also note the existence of a ferromagnetic region (F) at very small J as
predicted by high temperature expansions \cite{ferro_hightemp} or
ED \cite{ferro_ED}. Two crutial issues still remain to be addressed
namely the exact nature of the normal paramagnetic phase (P) and
possible pair-pair correlations (superconductivity) in the pair liquid phase.

\bigskip
I gratefully acknowledge many stimulating discussions with
M. Luchini and F. Mila.
{\it Laboratoire de Physique Quantique, Toulouse} is
{\it Unit\'e de Recherche Associ\'e au CNRS No 505}.
I also acknowledge support from the EEC Human Capital and Mobility
program under Grant No. CHRX-CT93-0332 and thank IDRIS (Orsay)
for allocation of CPU time on the C94 and C98 CRAY supercomputers.

%
%  Fig. 1
%
\begin{figure}
\caption{
Binding energy $\Delta_4$ vs J for clusters of 16, 20 and
26 sites. The 26-site cluster hole-hole binding energy $\Delta_2$
is also indicated by open stars.
}
\label{f1}
\end{figure}

%
%  Fig. 2
%
\begin{figure}
\caption{
Hole-hole correlation function for various hole separation vs J/t
obtained on a 26-site cluster with two (a) and four (b) holes.
The various symbols associated to the allowed distances are indicated on the
figure.
}
\label{f2}
\end{figure}

%
%  Fig. 3
%
\begin{figure}
\caption{
(a) energy per site vs hole density for 18 and 26 site clusters and
various J/t values (indicated on the plot). Continuous lines are
interpolations between data points corresponding to the same system sizes.
Extrapolation to the thermodynamic limit are indicated by dotted lines.
(b) data for J/t=1 only shown on an enlarged scale.
}
\label{f3}
\end{figure}

%
%  Fig. 4
%
\begin{figure}
\caption{
Static spin (a) and charge (b) structure factors $S(\bf q)$
and $N(\bf q)$ along symmetry lines of the
Brillouin zone for various J/t ratios as indicated on the plot.
$\Gamma$, M and X corresponds to $(0,0)$, $(\pi,\pi)$ and $(\pi,0)$
respectively. In (b) the dotted lines are obtained
assuming the $N_h$ holes behave as non-interacting spinless fermions
with various dispersion relations (see text).
}
\label{f4}
\end{figure}

%
%  Fig. 5
%
\begin{figure}
\caption{
Schematic picture of the J/t -- $n$ phase diagram.
}
\label{f5}
\end{figure}


\begin{references}
\bibitem{note1} The largest Hilbert space was realized
for 22 electrons on 26 sites. Full use of space group and spin inversion
symmetries was necessary to reduce the effective Hilbert space to
$50\, 717\, 244$ basis elements.
\bibitem{note} We cannot rule out PS at very small hole density and small J.
However, this regime is not relevant to us.
\end{references}
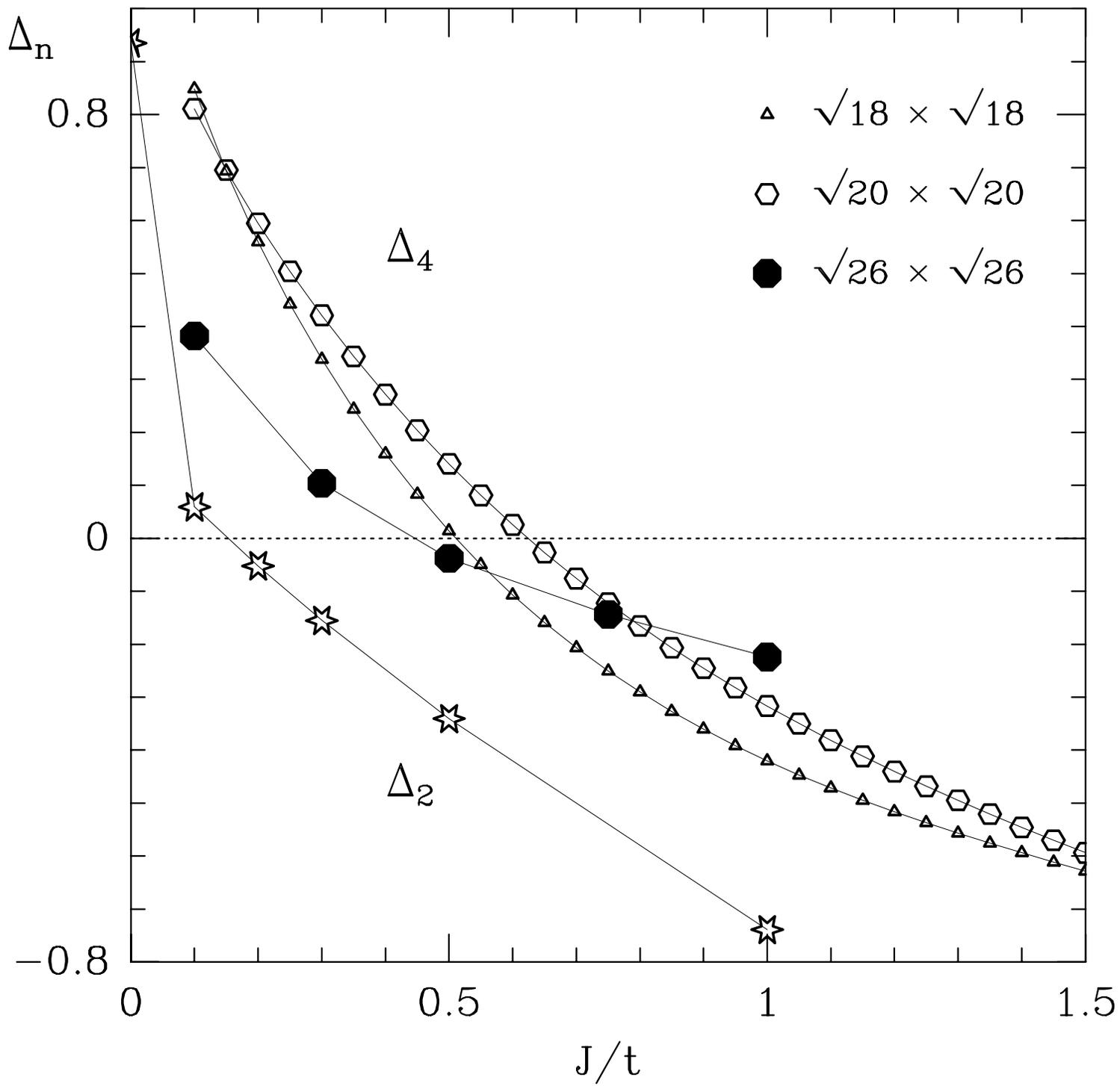
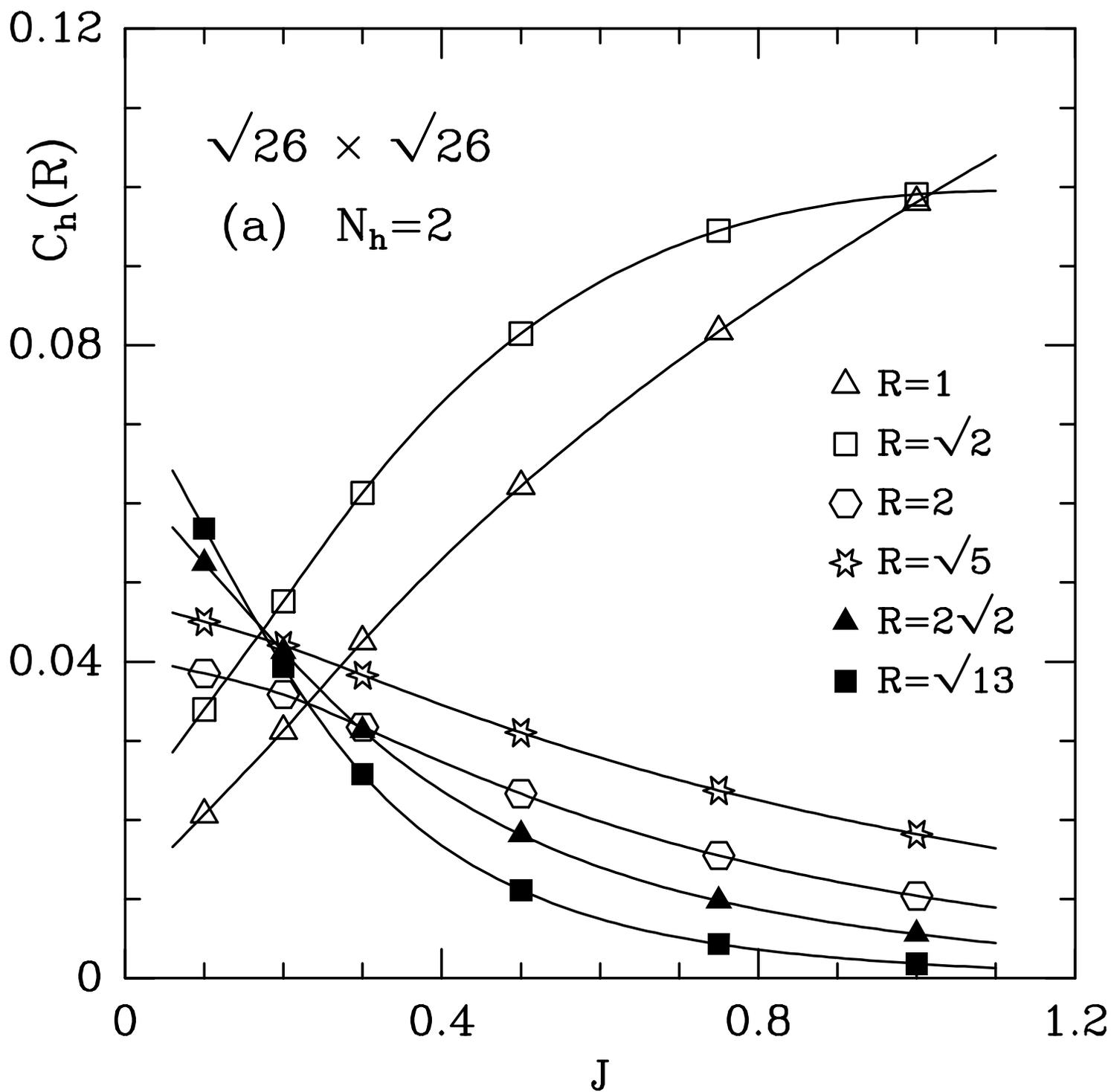
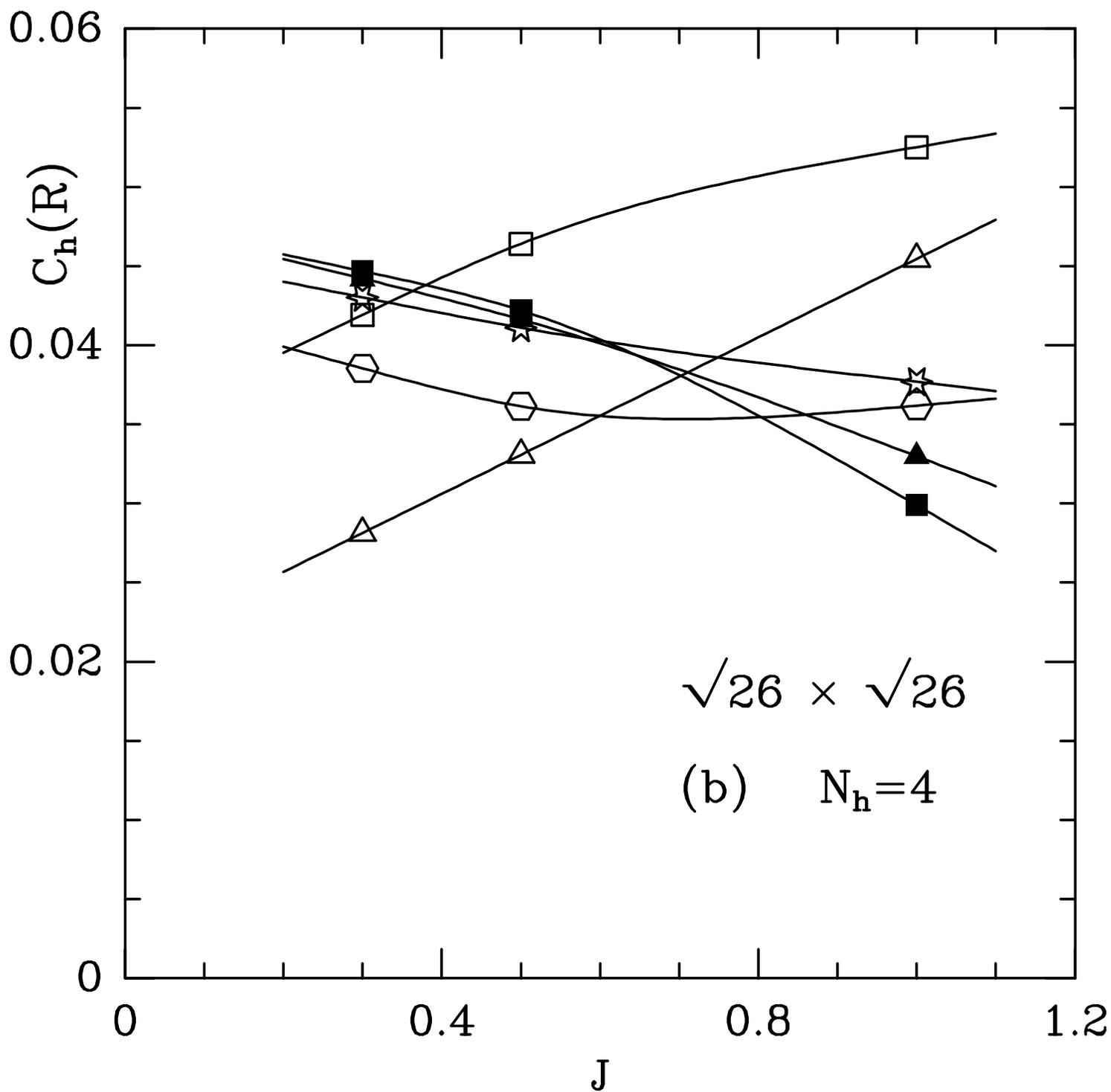
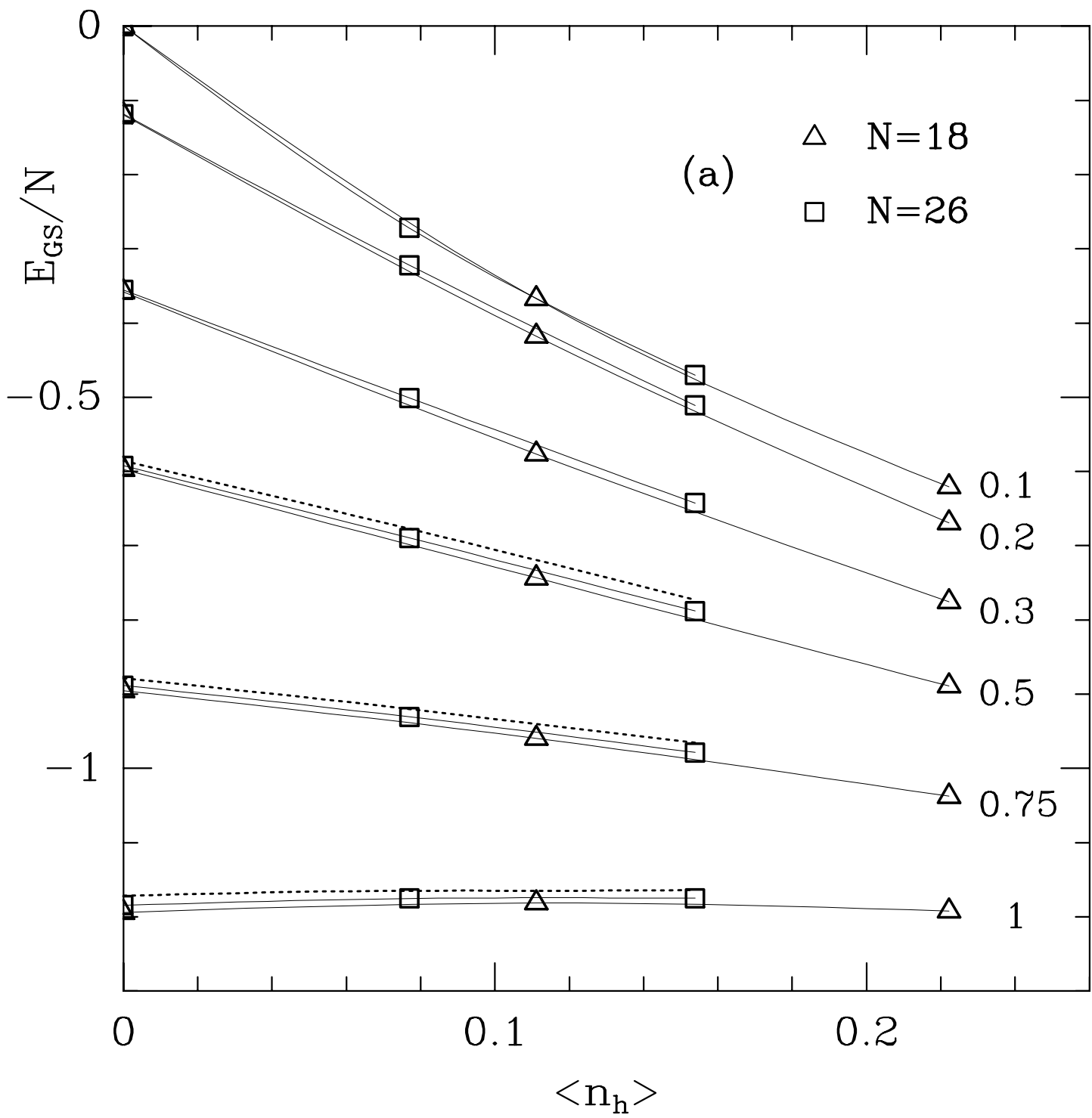
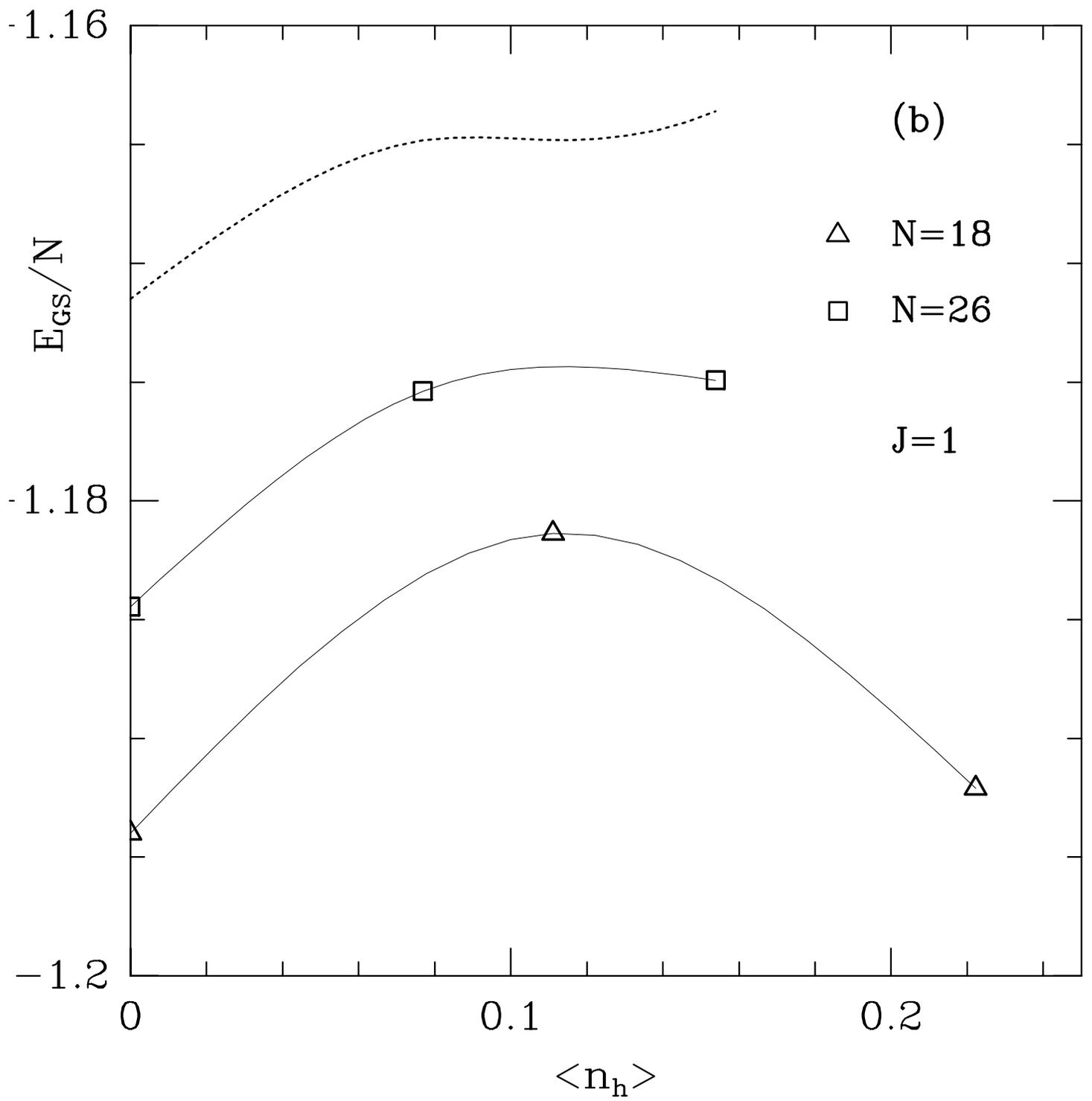
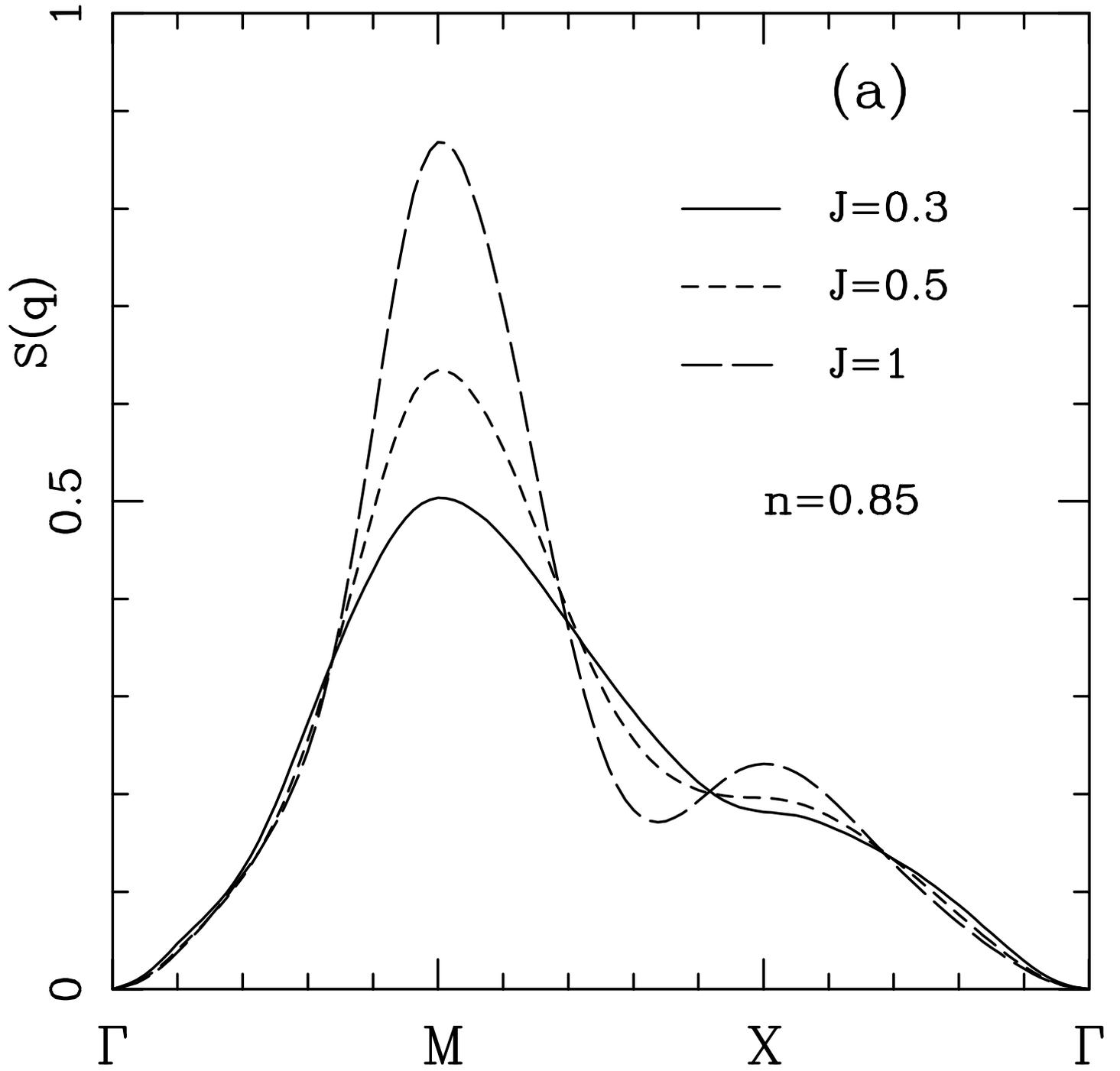
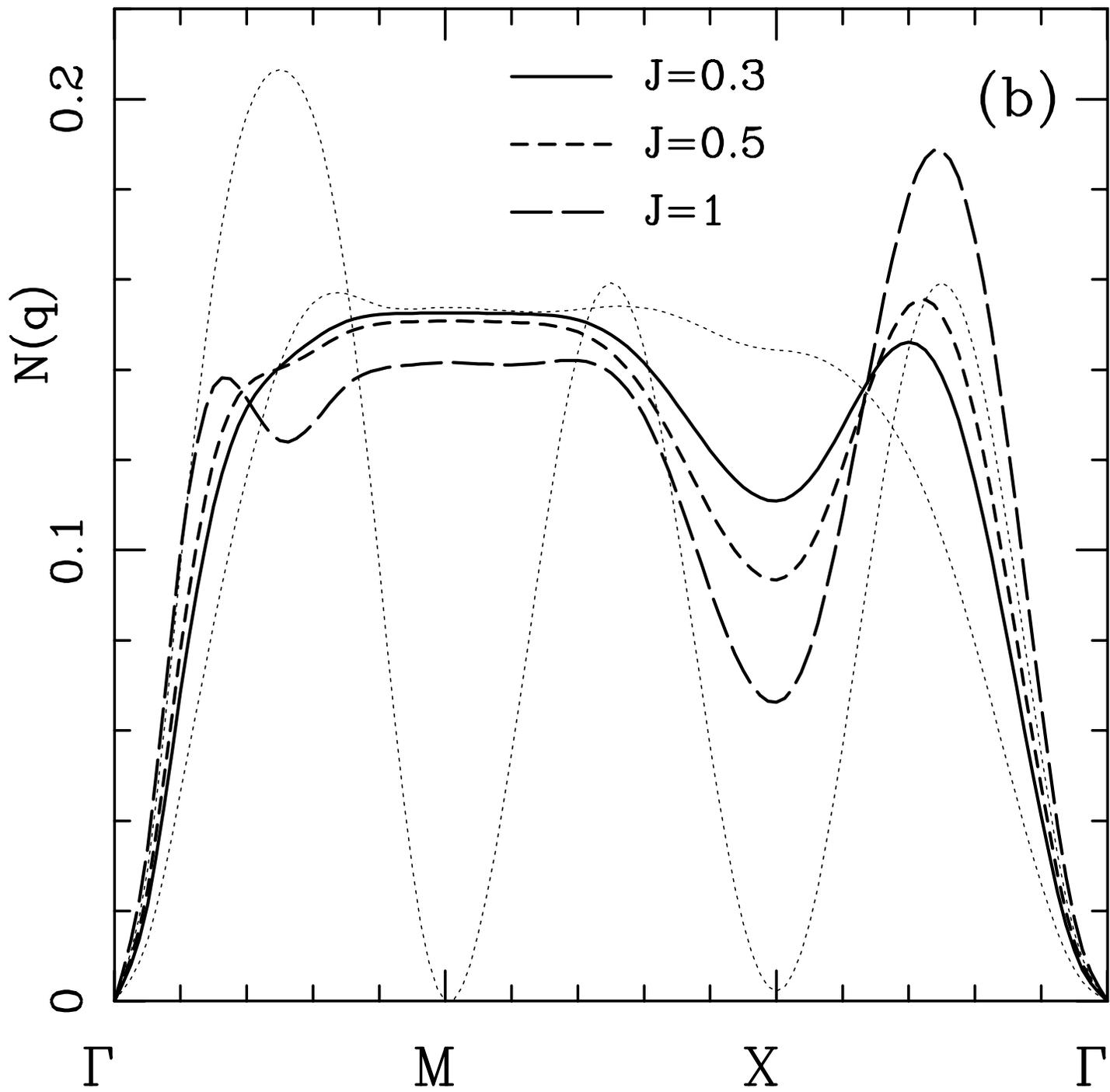
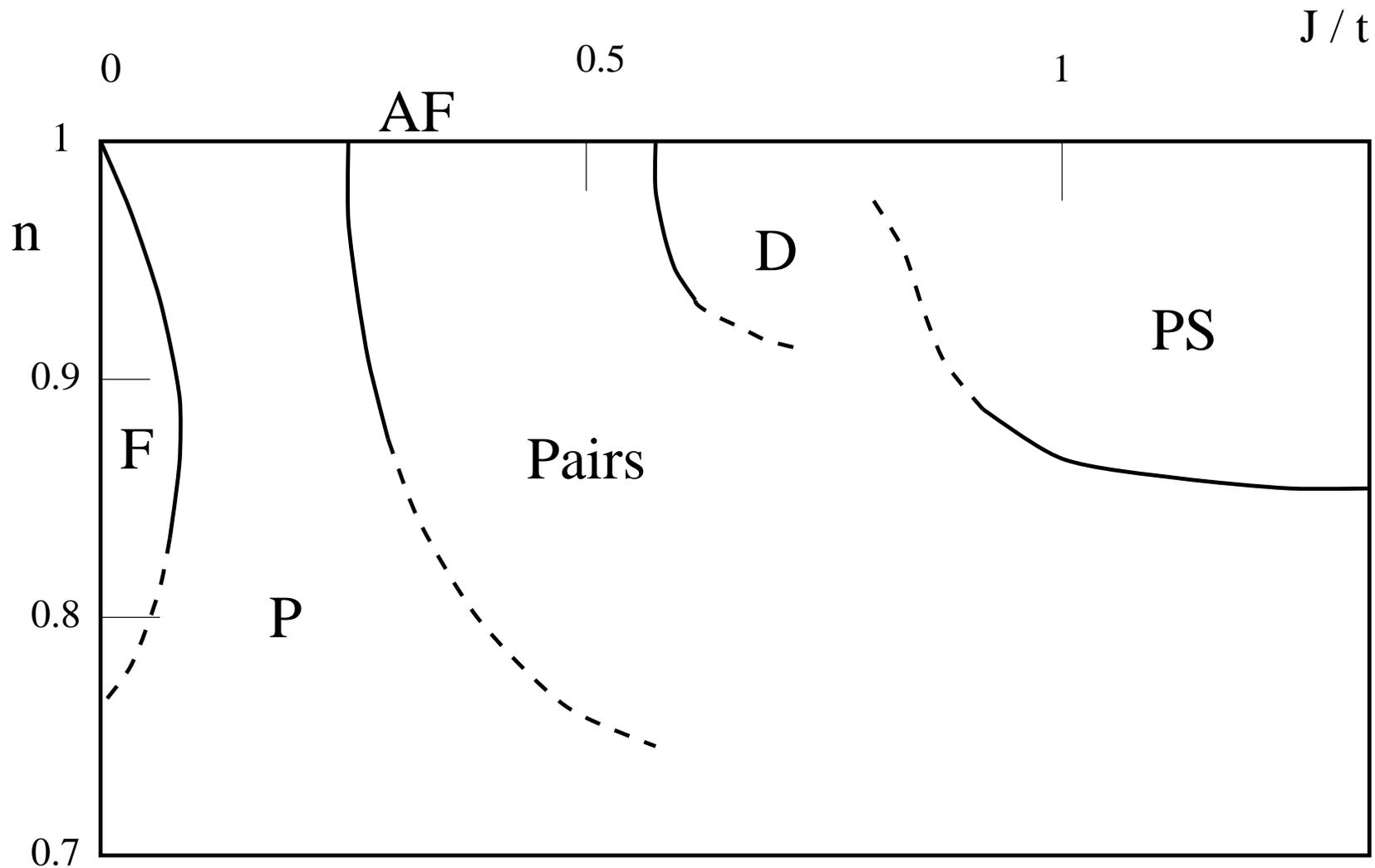
\end{document}